\documentclass[reprint, amsmath, amssymb, groupaddress, aps, prb]{revtex4-1}
\usepackage{graphicx}
\usepackage{amsmath}
\begin{document}
\title{Towards Obtaining 2D and 3D and 1D PtPN with Pentagonal Pattern}
\author{Duo Wang}
\author{Lei Liu}
\author{Houlong L. Zhuang}
\email{zhuanghl@asu.edu}
\affiliation{School for Engineering of Matter Transport and Energy, Arizona State University, Tempe, AZ 85287, USA}
\date{\today}
\begin{abstract}
We apply an alloying strategy to single-layer PtN$_2$ and PtP$_2$, aiming to obtain a single-layer Pt-P-N alloy with a relatively low formation energy with reference to its bulk structure. We perform structure search based on a cluster-expansion method and predict single-layer and bulk PtPN consisting of pentagonal networks. The formation energy of single-layer PtPN is significantly lower in comparison with that of single-layer PtP$_2$. The predicted bulk structure of PtPN adopts a structure that is similar to the pyrite structure. We also find that single-layer pentagonal PtPN, unlike PtN$_2$ and PtP$_2$, exhibits a sizable, direct PBE band gap of 0.84 eV. Furthermore, the band gap of single-layer pentagonal PtPN calculated with the hybrid density functional theory is 1.60 eV, which is within visible light spectrum and promising for optoelectronics applications. In addition to predicting PtPN in the 2D and 3D forms, we study the flexural rigidity and electronic structure of PtPN in the nanotube form. We find that single-layer PtPN has similar flexural rigidity to that of single-layer carbon and boron nitride nanosheets and that the band gaps of PtPN nanotubes depend on their radii. Our work shed light on obtaining an isolated 2D planar, pentagonal PtPN nanosheet from its 3D counterpart and on obtaining 1D nanotubes with tunable bandgaps. 
\end{abstract}
\maketitle
\section{Introduction}
Two-dimensional (2D) materials such as single-layer graphene and boron nitride hold great promise for a wide range of applications such as electronic devices,\cite{farmer2010graphene,eda2008large} optoelectronic systems,\cite{wu2008organic,dean2009second,xia2009ultrafast} and energy-related applications.\cite{wang2013interconnected,liu2018porous,raccichini2015role} Hexagons dominate the building blocks of many 2D materials, which have issues such as the absence of anisotropy, a desirable feature for certain applications like photodetectors.\cite{li2018highly} To introduce anisotropy, one may resort to 2D materials that adopt other shapes\textemdash in particular, pentagons\textemdash as their building blocks. 

Because only 15 types of pentagons can tessellate an infinite plane and pentagons possess intrinsic anisotropy,\cite{rao2017exhaustive} 2D materials consisting of a pattern of pentagons represent an important addition to the large family of 2D materials whose structures are dominated by patterns of other shapes especially hexagons. As two most promising examples, single-layer PtP$_2$ and PtN$_2$\cite{liu2018penta,liu2018ptp,yuan2019single,zhao20192d} have been predicted to exhibit a unique planar, pentagonal structure and attractive electronic structures such as direct band gaps calculated at the level of hybrid density functional theory\textemdash note that the band gaps at the level of Perdew--Burke--Ernzerhof (PBE) functional theory are negligibly small.\cite{liu2019dimension,liu2018ptp} But the stability of the bulk counterparts of these two single-layer pentagonal materials and their formation energies are likely to prohibit successful synthesis or exfoliation. In particular, bulk PtN$_2$ with the pyrite structure is stable only at high pressures.\cite{crowhurst2006synthesis} As a result, the formation energy of single-layer PtN$_2$ is unphysically negative (i.e., energy is gained from reducing bulk to single-layer PtN$_2$) if using the pyrite structure as the reference. On the other hand, although bulk PtP$_2$ crystallizes as the pyrite structure at ambient conditions,\cite{Thomassen1929} the theoretical formation energy of single-layer PtP$_2$ could be too high (positive) to exist as an isolated nanosheet. 

In this work, we apply density functional theory (DFT) calculations and a cluster expansion method to search for stable single-layer (2D) and bulk (3D) Pt-P-N alloys based on single-layer PtN$_2$ and PtP$_2$ by taking the advantage of the low formation energy of single-layer PtN$_2$ and stable bulk counterpart of single-layer PtP$_2$. In addition to designing 2D and 3D Pt-P-N alloys, we also examine the feasibility of obtaining 1D Pt-P-N nanotubes from bending 2D Pt-P-N nanosheets, with the goal of achieving tunable electronic structures. 
\section{Methods}
We perform the DFT calculations with the Vienna {\it Ab-initio} Simulation Package (VASP, version 5.4.4).\cite{Kresse96p11169} We use the PBE functional for describing the exchange-correlation interactions.\cite{Perdew96p3865} We also use the standard Pt, P, and N potential datasets based on the PBE functional along with the projector-augmented wave (PAW) method.\cite{Bloechl94p17953,Kresse99p1758} Among the potentials, the 5$d^9$ and 6$s$ electrons of Pt atoms, the 3$s^2$ and 3$p^3$ electrons of P atoms, and the 2$s^2$ and 2$p^3$ electrons of N atoms are treated as valence electrons. We adopt the plane waves with the cut-off kinetic energy of 550 eV to approximate the electron wave functions. We use $\Gamma$-centered $12~\times~12~\times~12$, $12~\times~12~\times~1$, and $9~\times~1~\times~1$ Monkhorst-Pack\cite{PhysRevB.13.5188} $k$-point grids to sample the $k$ points in the reciprocal space for 3D, 2D, and 1D PtPN, respectively. A vacuum spacing of 18.0~\AA~is applied to the surface slabs simulating isolated 2D and 1D PtPN. For 3D, 2D, and 1D PtPN, we optimize the atomic positions completely. Furthermore, for 3D PtPN, we relax all the lattice parameters; for 2D PtPN, we optimize the in-plane lattice constants; for 1D PtPN, we optimize the cell length along the axial direction. The force criterion for all of the geometry-optimization calculations is set to 0.01 eV/\AA.
\section{Results and Discussion}
\begin{figure}
  \includegraphics[width=8cm]{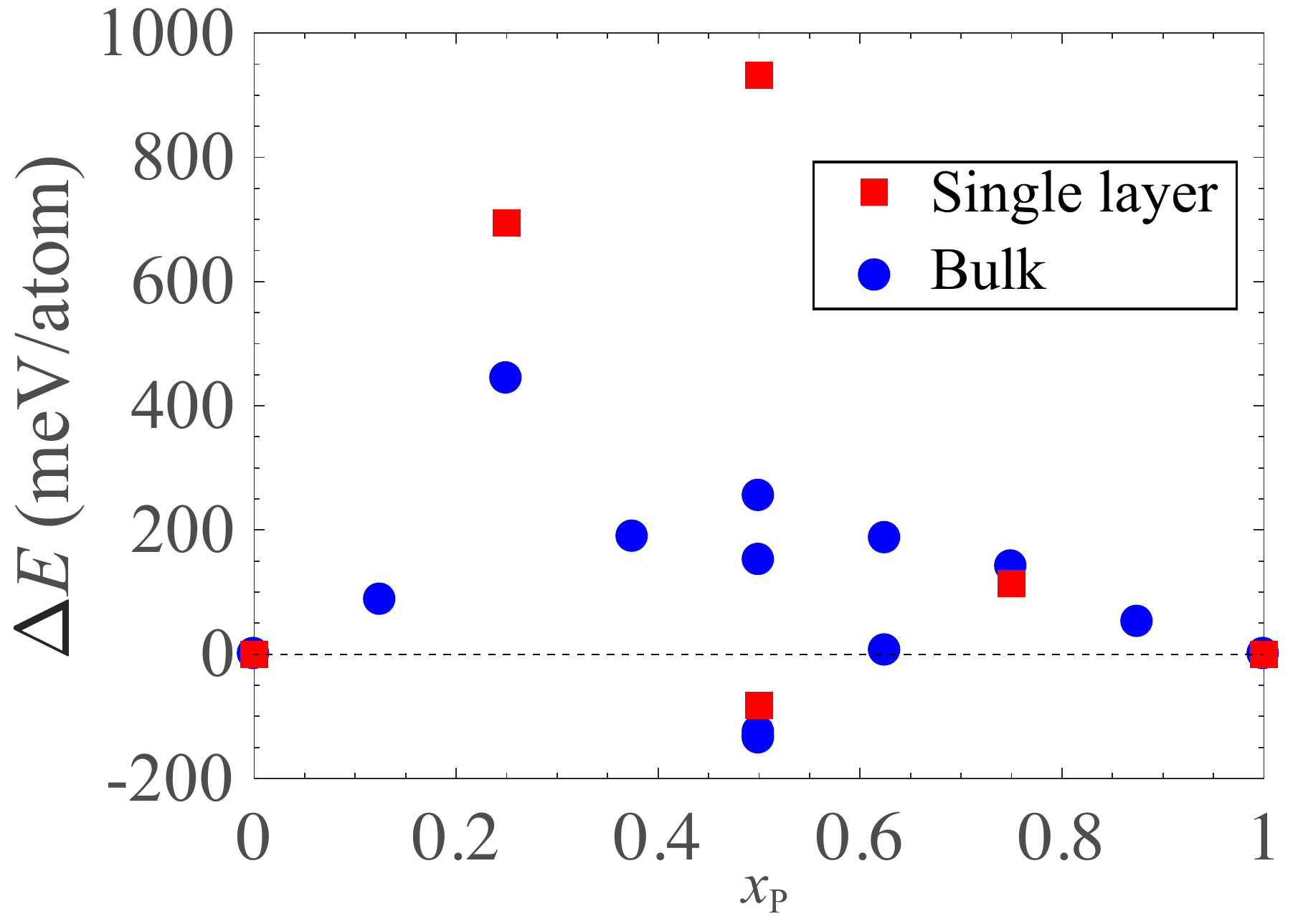}
  \caption{Energy difference $\Delta E$ as a function of the concentration of P $x_\mathrm{P}$ in single-layer and bulk PtPN with chemical formulas $\mathrm{Pt_2N}_{4(1-x)}\mathrm{P}_{4x}$ and $\mathrm{Pt_4N}_{8(1-x)}\mathrm{P}_{8x}$, respectively.}
  \label{fig:ediff}
\end{figure}

\begin{figure}
  \includegraphics[width=8cm]{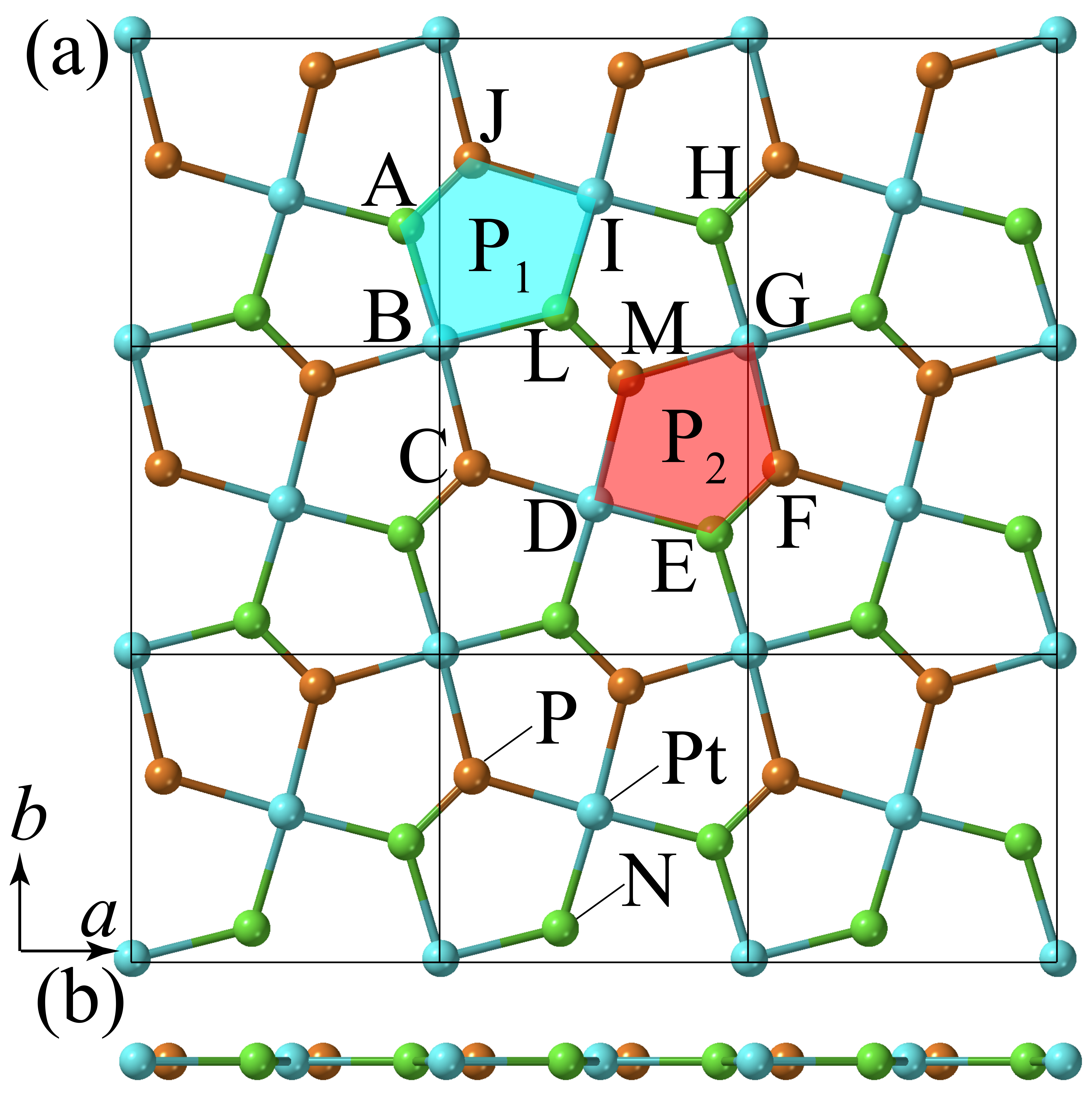}
  \caption{(a) Top and (b) side views of a $3\times 3 \times 1$ supercell of single-layer PtPN. Two distinct pentagons denoted as P$_1$ and P$_2$ are enclosed in the cyan and red shaded areas.}
  \label{fig:structure}
\end{figure}

We first use the Alloy Theoretic Automated Toolkit (ATAT) to generate inequivalent structures of single-layer $\mathrm{Pt_2N}_{4(1-x)}\mathrm{P}_{4x}$ with five different concentrations $x$ of P ($x$ = 0, 0.25, 0.5, 0.75, and 1) and to automate the geometry optimizations and energy calculations.\cite{avdw:atat} The chemical formula written in this form is because each unit cell of single-layer PtN$_2$ and PtP$_2$ consists of two Pt atoms and four N/Pt atoms (each Pt atom is four-fold coordinated by N/P atoms; each N/P atom is three-fold coordinated by the same atoms). When $x$ = 0 and 1, the systems correspond to single-layer PtN$_2$ and PtP$_2$, respectively. For $x$ = 0.25, one P atom can replace any of the four N atoms in a unit cell, but all of the four structures are equivalent due to the four-fold rotational symmetry in single-layer PtN$_2$ and PtP$_2$. Therefore, only one of these four structures is optimized and its energy is calculated. For $x$ = 0.5, there are two different structures with and without a center of inversion symmetry, respectively. Similarly, for $x$ = 0.75, there is only one inequivalent structure. We compute the energy change $\Delta E$ of the following `reaction': 
\begin{equation}
(1-x)\mathrm{Pt_2N_4} + x \mathrm{Pt_2P_4} \rightarrow  \mathrm{Pt_2N}_{4(1-x)}\mathrm{P}_{4x}.
\label{eq00}
\end{equation}
By this definition $\Delta E$ = 0, when $x$ = 0 or 1. Figure~\ref{fig:ediff} shows the $\Delta E$ results for the six single-layer $\mathrm{Pt_2N}_{4(1-x)}\mathrm{P}_{4x}$ structures optimized from VASP calculations. We find that the structure with $x$ = 0.5 is the most stable, corresponding to the chemical formula PtPN. The energy difference between single-layer PtPN and PtN$_2$ and PtP$_2$, i.e., $\Delta E$ = $E_\mathrm{PtPN}$-($E_\mathrm{PtN_2}$+$E_\mathrm{PtP_2}$)/2 is determined as $\Delta E$ = -82 meV/atom. Figure~\ref{fig:structure} shows the top and side views of a $3 \times 3 \times 1$ supercell of this structure. Similar to single-layer PtN$_2$ and PtP$_2$, single-layer pentagonal PtPN exhibits a completely planar structure and the optimized in-plane lattice constants $a$ and $b$ are 5.30 and 5.29~\AA, respectively.

The optimized structure of single-layer PtPN exhibits no four-fold rotational symmetry, so the two in-plane lattice constants are not identical. As a result, instead of observing a pattern of same pentagons as in single-layer PtN$_2$ and PtP$_2$\textemdash{these pentagons below to the same type and the tessellation pattern from this type of pentagons is called the Cairo tessellation}, there are two different types of pentagons in single-layer PtPN. Table~\ref{summary2} lists the side lengths (bond lengths) and angles (bond angles) forming the two distinct pentagons illustrated in Fig.\ref{fig:structure}(a). Referring to the definitions for the 15 types of pentagons that can monohedrally tile a plane,\cite{mann2018convex} neither of the two types of pentagons in single-layer PtPN belongs to any of the 15 types. Therefore, the geometry of PtPN shows an example that a plane can still be tiled gaplessly by a combination of different types of pentagons from the 15 ones, retaining the anisotropy for a 2D material. 

\begin{figure}
  \includegraphics[width=8cm]{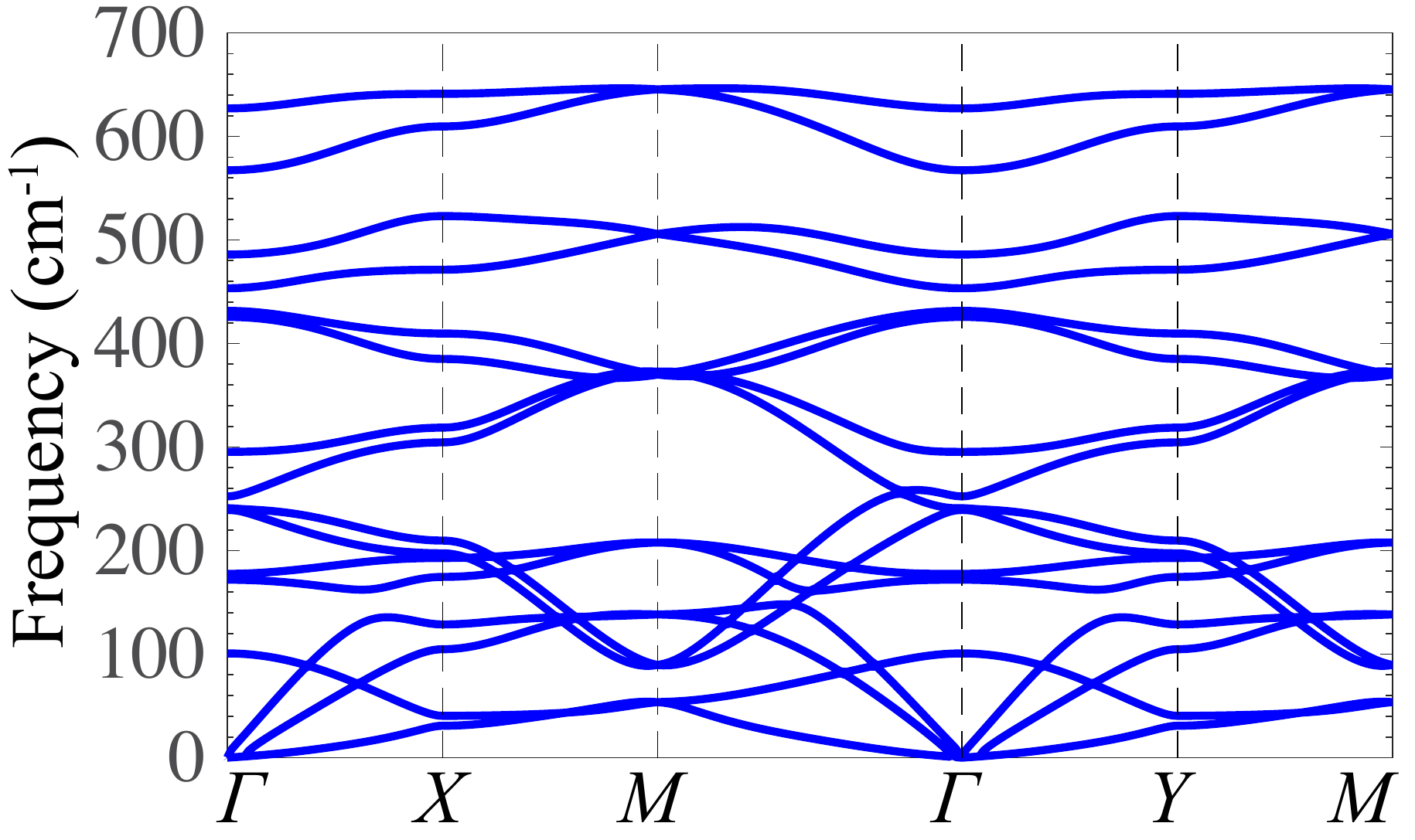}
  \caption{Predicted phonon spectrum of single-layer PtPN.}
  \label{fig:ph}
\end{figure}

We next examine the dynamical stability of the predicted structure of single-layer PtPN. Figure~\ref{fig:ph} shows the computed phonon spectrum of single-layer PtPN. The real phonon frequencies confirm the dynamical stability of the completely planar structure of single-layer PtPN.


\begin{figure}
  \includegraphics[width=8cm]{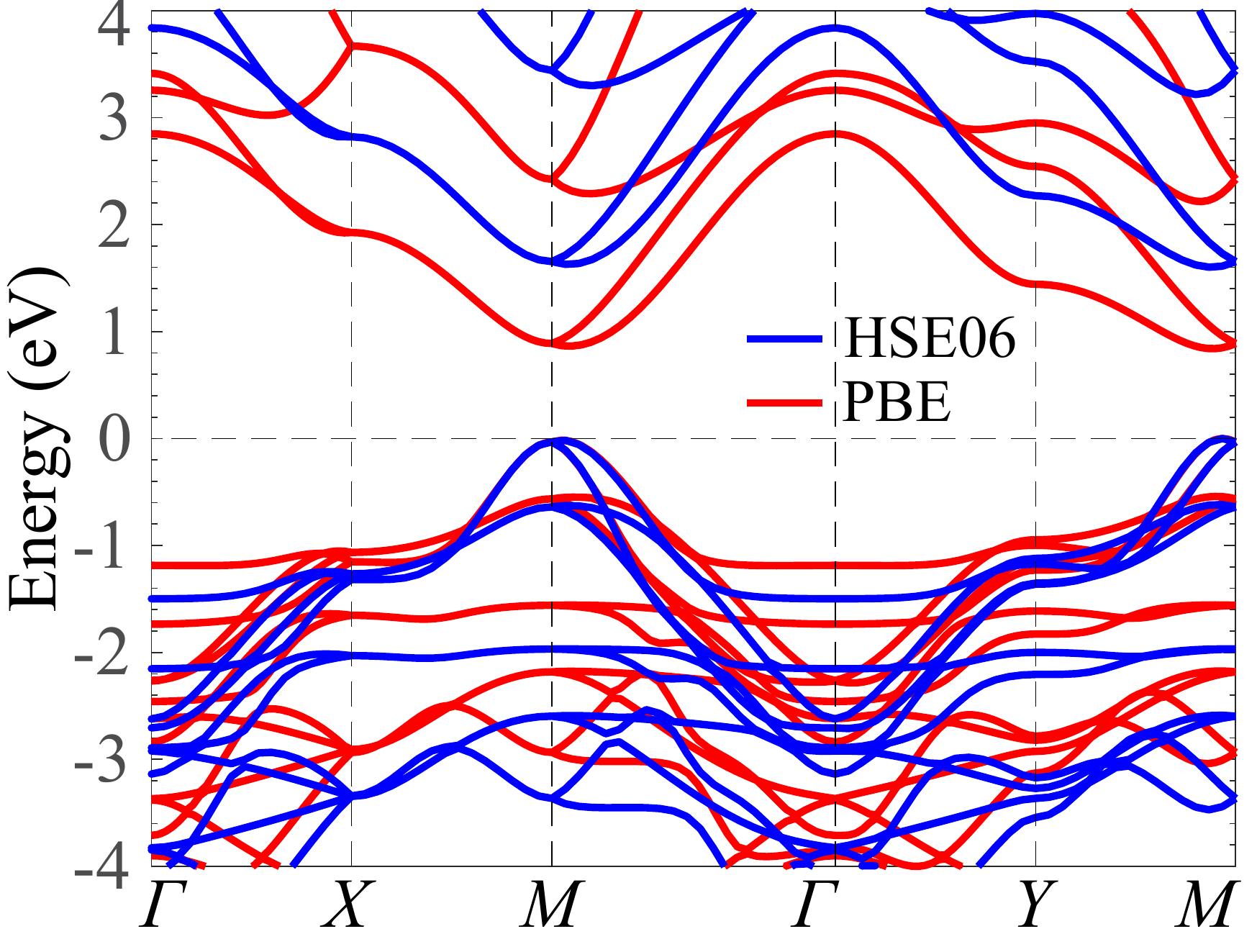}
  \caption{Band structures of single-layer PtPN calculated with the PBE and HSE06 functionals.}
  \label{fig:bandstructure}
\end{figure}

\begin{figure}
  \includegraphics[width=8cm]{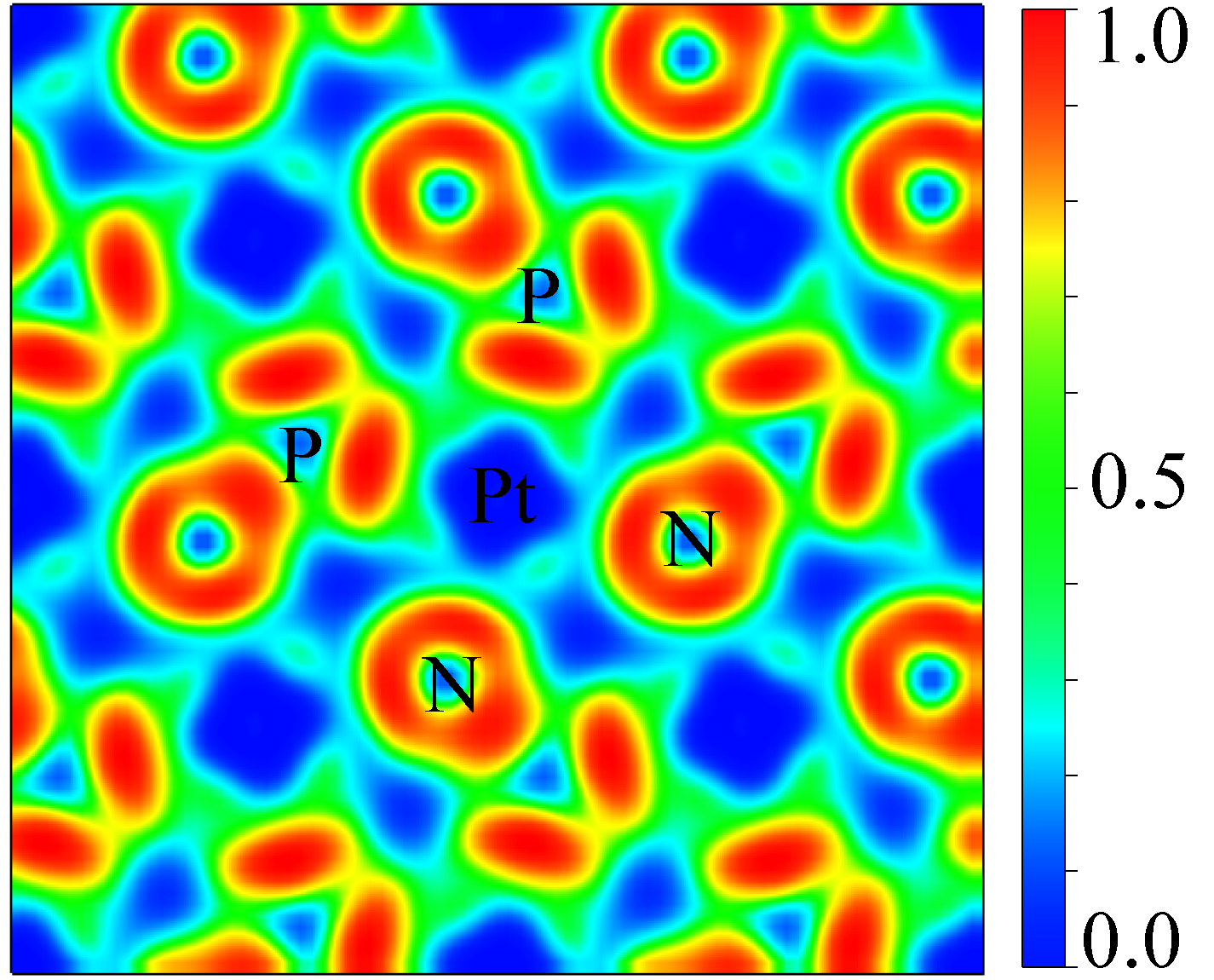}
  \caption{Electron localization function of single-layer PtPN.}
  \label{fig:elf}
\end{figure}

\begin{table*}[!htbp]
\caption{Bond lengths (in \AA) and angles (in degrees) of the two distinct pentagons P$_1$ and P$_2$ embedded in the atomic structure of single-layer pentagonal PtPN (see Fig.\ref{fig:structure}(a)).}
  \begin{ruledtabular}
    \begin{center}
      \begin{tabular}{ccccccccccc}
      Pentagon&AB&BL&LI&IJ&JA&$\angle$ABL&$\angle$BLI&$\angle$LIJ&$\angle$IJA&$\angle$JAB\\
      \hline
     P1& 2.09&2.12&2.09&2.20&1.60&92.47&120.80&89.54&118.88&118.32\\
      \hline
      &ED&DM&MG&GF&FE&$\angle$EDM&$\angle$DMG&$\angle$MGF&$\angle$GFE&$\angle$FED\\
            \hline
      P2 & 2.12 & 2.22 & 2.20 & 2.22 & 1.60 & 90.21 & 120.13 & 87.78 & 121.00 & 120.88\\
       \end{tabular}
 \end{center}
 \end{ruledtabular}
 \label{summary1}
\end{table*}
Figure~\ref{fig:bandstructure} shows the band structures of single-layer PtPN calculated with the PBE and Heyd-Scuseria-Ernzerhof (HSE06)\cite{heyd2003hybride} functionals. Unlike single-layer PtN$_2$ and PtP$_2$, where the PBE band gaps are nearly zero,\cite{liu2018penta,liu2018ptp,yuan2019single,zhao20192d} the PBE band gap of single-layer PtPN has already shown a direct band gap of 0.84 eV. The conduction band minimum and valence band maximum both locate at a $k$ point near the $M$ point. Using the HSE06 hybrid density functional theory corrects the band gap to 1.60 eV, much larger than the HSE06 band gaps of single-layer pentagonal PtN$_2$ (1.11 eV)\cite{liu2019dimension} and PtP$_2$ (0.52 eV).\cite{PhysRevMaterials.2.114003} The HSE06 band gap indicates that single-layer PtPN is promising for optoelectronics applications that can utilize the direct band gap within visible light spectrum. Comparing with single-layer PtN$_2$ and PtP$_2$, the enhanced band gaps in single-layer PtPN may be correlated with the bonding characteristics, which can be revealed from the electron localization function (ELF) shown in  Fig.~\ref{fig:elf}. Different from single-layer PtN$_2$ and PtP$_2$ showing both ionic and covalent bonding characteristics,\cite{liu2018penta,liu2018ptp,yuan2019single,zhao20192d} the bonding type in single-layer PtPN is dominantly ionic and ionic bond (e.g., in BN) is often associated with large band gaps due to the large charge transfer between cations and anions.

\begin{figure}
  \includegraphics[width=8cm]{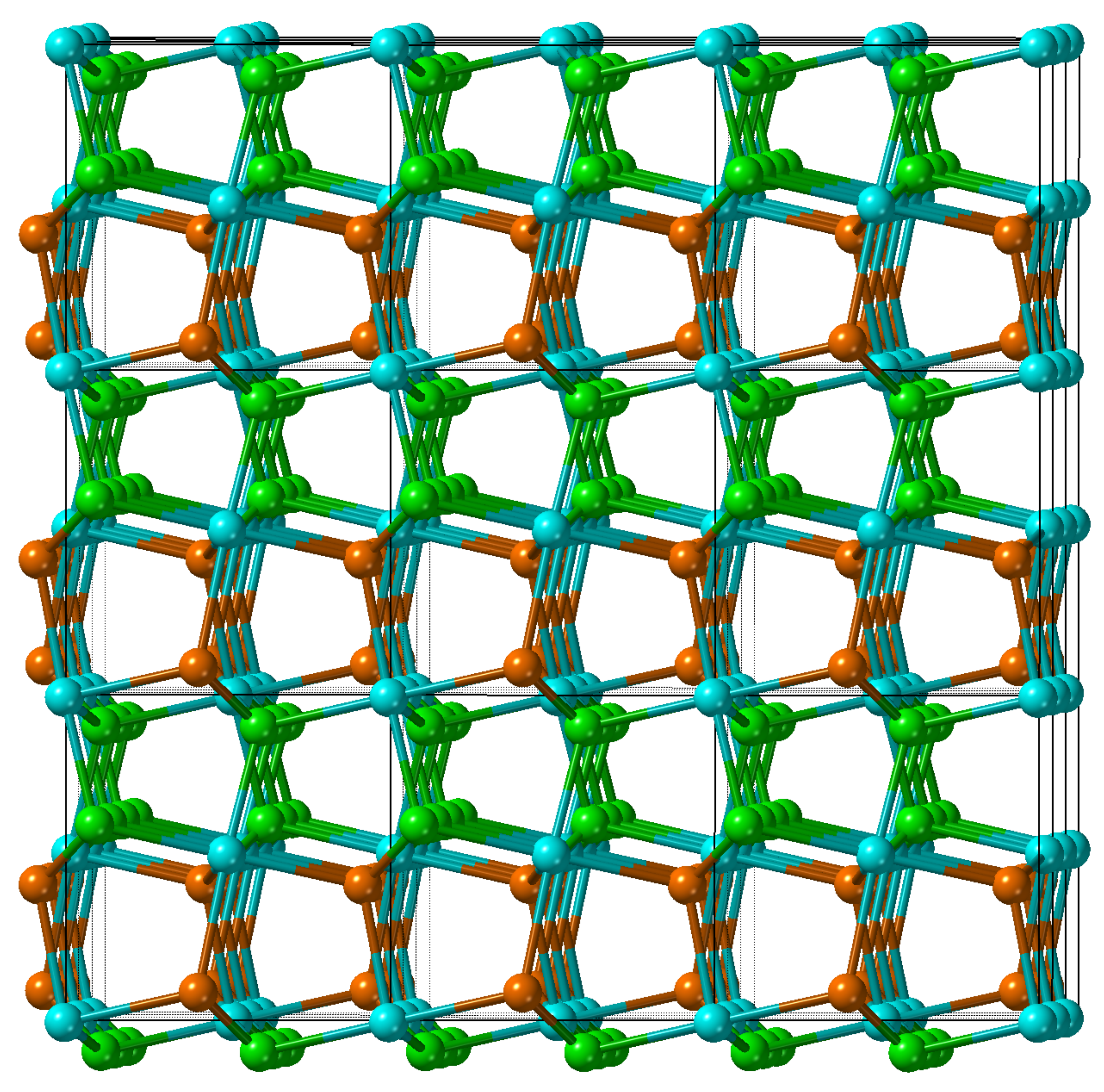}
  \caption{A $3\times 3 \times 3$ supercell of the predicted bulk structure of PtPN.}
  \label{fig:3d}
\end{figure}

Having predicted the stable structure of single-layer PtPN and its attractive direct band gap, we aim to predict the bulk counterpart from which single-layer PtPN could be exfoliated. The existence of a bulk counterpart appears to be a necessary condition for all the 2D materials that have been successfully synthesized or exfoliated. We apply ATAT again to enumerate all the possible bulk structures at different concentrations of P for bulk $\mathrm{Pt_4N}_{8(1-x)}\mathrm{P}_{8x}$ in a 12-atom unit cell. We compute the $\Delta E$ for the following `reaction': 
\begin{equation}
(1-x) \mathrm{Pt_4N_8} + x \mathrm{Pt_4P_8} \rightarrow  \mathrm{Pt_4N}_{8(1-x)}\mathrm{P}_{8x}.
\label{eq0}
\end{equation}
Figure~\ref{fig:ediff} display all the $\Delta E$ results. Similar to the single-layer cases, the most stable bulk compound occurs at $x$ = 0.5, and the structure is illustrated in Fig.~\ref{fig:3d}. This bulk structure is nearly a cubic structure with space group $P_{ca}$2$_1$ and the lattice constants are 5.301, 5.301, and 5.305~\AA, respectively. Moreover, the bulk structure resembles the pyrite structure adopted by bulk PtP$_2$.\cite{Thomassen1929} Namely, viewing along the $a/b/c$ axis, the bulk structure can be regarded stacked PtPN single layers with a buckled structure. In contrast to PtP$_2$, the stable bulk structure of PtN$_2$ remains unclear. If assuming bulk PtN$_2$ also adopts the pyrite structure, we encounter an incorrect conclusion that single-layer PtN$_2$ is more stable than bulk PtN$_2$ with the pyrite structure.\cite{liu2019dimension} We recently proposed a new structure of bulk PtN$_2$ with layered structure. We also compute the energy of bulk PtPN with a similar layered structure, but it is higher than that of bulk PtPN with the pyrite-type structure by 180 meV/atom, confirming that the latter structure is the most stable one.

With the predicted stable structure of bulk PtPN, we calculate the formation energy of single-layer PtPN, namely the energy difference between single-layer PtPN and the stable bulk structure. We find that the formation energy is 174 meV/atom, which is much smaller than that (410 meV/atom calculated with the PBE functional)\cite{liu2018ptp} of single-layer pentagonal PtP$_2$. The formation energy of a 2D material at this scale often implies the 2D material could be exfoliated if the bulk counterpart exists or synthesized if there is no bulk counterpart.\cite{singh2015computational} The small theoretical formation energy of single-layer PtPN suggests a possible route to obtain this single-layer material, i.e., alloying the stable bulk PtP$_2$ compound by N atoms to obtain the bulk structure of PtPN and then applying the mechanical exfoliation method to the ternary bulk compound to acquire single-layer sheets of PtPN. Alternatively, it is also worth attempting the molecular beam epitaxy method\cite{fu2017molecular} to obtain the single-layer sheets.   

\begin{figure}
  \includegraphics[width=8cm]{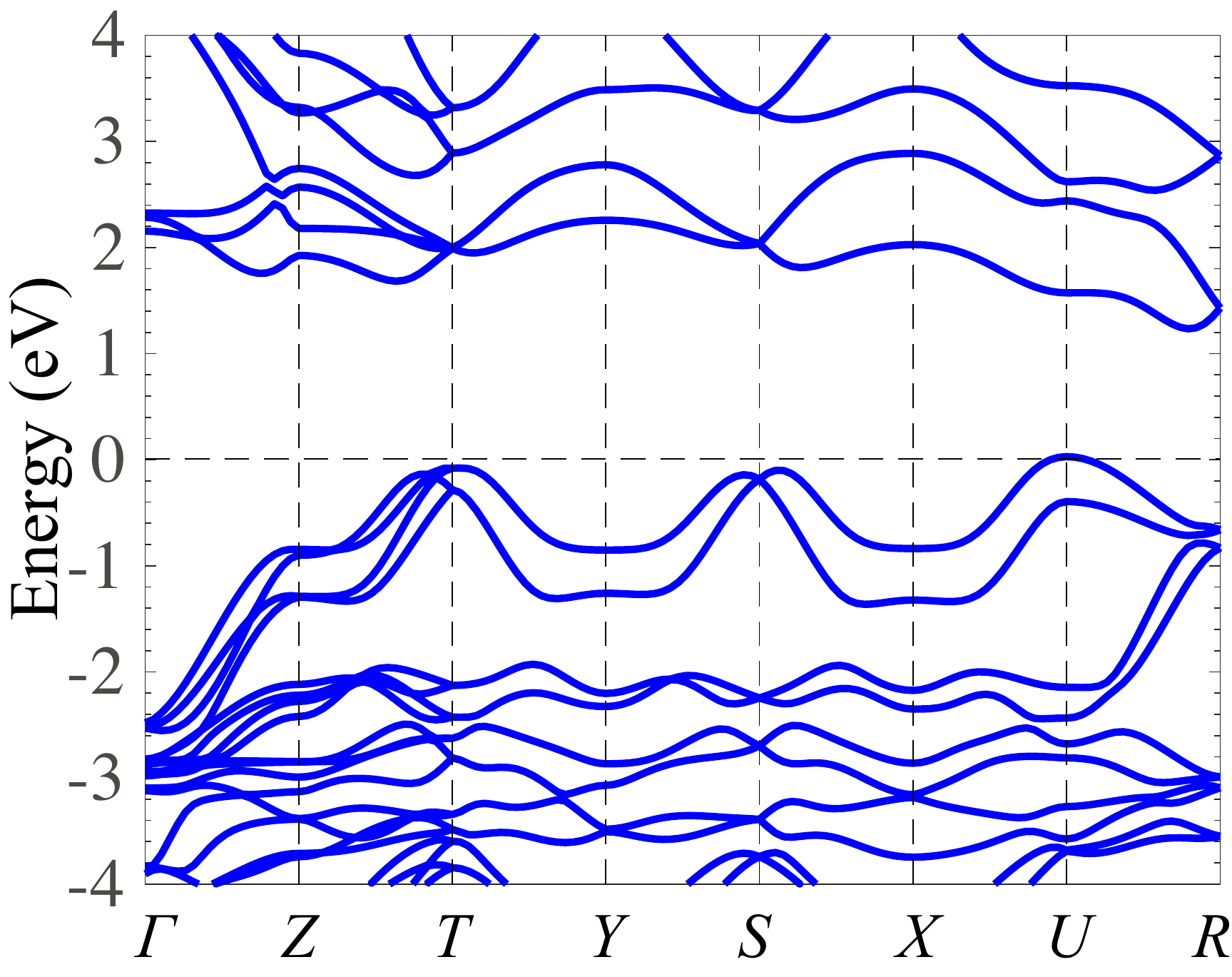}
  \caption{Band structure of bulk PtPN calculated with the PBE functional. The fractional coordinates of the high-symmetry $k$-points are {\it{$\Gamma$}} (0, 0, 0), {\it Z} (0, 0, 1/2), {\it T} (0, 1/2, 1/2), {\it Y} (0, 1/2, 0), {\it S} (1/2, 1/2, 0), {\it X} (1/2, 0, 0), {\it U} (1/2, 0, 1/2), {\it R} (1/2, 1/2, 1/2).}
  \label{fig:bulkband}
\end{figure}

Figure~\ref{fig:bulkband} shows the PBE band structure of bulk PtPN with the pyrite-type structure. As can be seen, it is semiconducting with an indirect band gap of 1.21 eV. Bulk PtN$_2$ and PtP$_2$ with the pyrite structure are also found to have indirect PBE band gaps of 1.35 \cite{liu2019dimension} and 1.06 eV,\cite{PhysRevMaterials.2.114003} respectively. It seems to be expected that the band gap of bulk PtPN lies between those of bulk PtN$_2$ and PtP$_2$. This trend also holds as the HSE06 band gaps of bulk PtN$_2$, PtPN, and PtP$_2$ are 2.22,\cite{liu2019dimension} 1.76, and 1.59 eV,\cite{PhysRevMaterials.2.114003} respectively. For the three materials, one common feature is the decrease in their band gaps due to the dimension reduction\textemdash the PBE band gaps of single-layer PtN$_2$ and PtP$_2$ are so small that they should probably be regarded as metallic.\cite{liu2019dimension}  The PBE functional therefore not only underestimates the band gap of single-layer PtPN, but also performs poorly in determining the electronic structures (metallic or semiconducting) of single-layer PtN$_2$ and PtP$_2$ possibly due to their four-fold rotational symmetry, leading to the degenerate energy levels at the $M$ point and at the Fermi level. By contrast, all the single-layer forms of these three systems have direct band gaps at the HSE06 hybrid density functional level of theory and the HSE06 band gap of single-layer PtPN no longer lies between those of single-layer PtN$_2$ and PtP$_2$.

\begin{figure}
  \includegraphics[width=8cm]{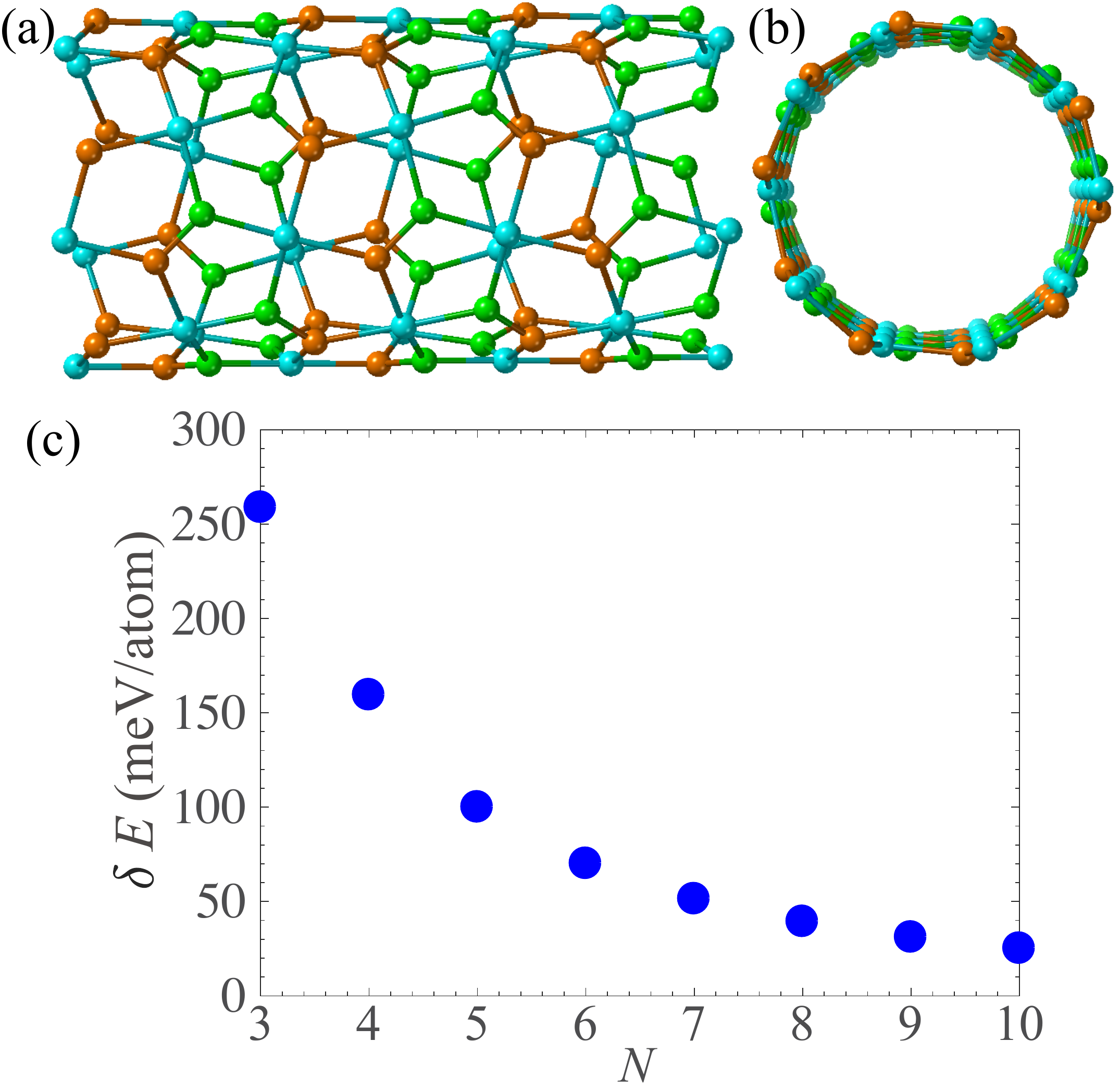}
  \caption{(a) Side and (b) top views of a PtPN nanotube. (c) Energy difference of PtPN nanotubes with reference to single-layer PtPN as a function of $N$, which determines the number of unit cells of single-layer PtPN that form a nanotube.}
  \label{fig:nts}
\end{figure}

If 3D and 2D PtPN can be obtained, one naturally continues to explore the structures and properties of 1D PtPN, i.e., PtPN nanotubes. We create the simulation models of PtPN nanotubes by bending a $N$ $\times$ 1 $\times$ 1 ($N$ ranges from 3 to 10) supercell of single-layer PtPN about the $b$ axis into a tube. The integer $N$ therefore determines the radius $R$ of a PtPN nanotube. Because the unit cell of single-layer PtPN is nearly in a square shape, bending about the $a$ axis results in nearly the same PtPN nanotubes.  We first assess the energy change $\delta E$ from 2D to 1D PtPN as a function of $N$. $\delta E$ stands for the average energy change between a flat atomic layer and an nanotube due to the change in curvature, which represents the energy required per atom to bend the flat PtPN nanosheet into the PtPN nanotubes of different radii.\cite{cherian2007elastic}  Figure~\ref{fig:nts} shows that $\delta E$ decreases as $N$ increases and the decrease is more significant when the $N$ values are small. As $N$ is close to infinity, the nanotubes are similar to 
single-layer sheets and $\delta E$ therefore approaches to zero.

\begin{figure}
  \includegraphics[width=8cm]{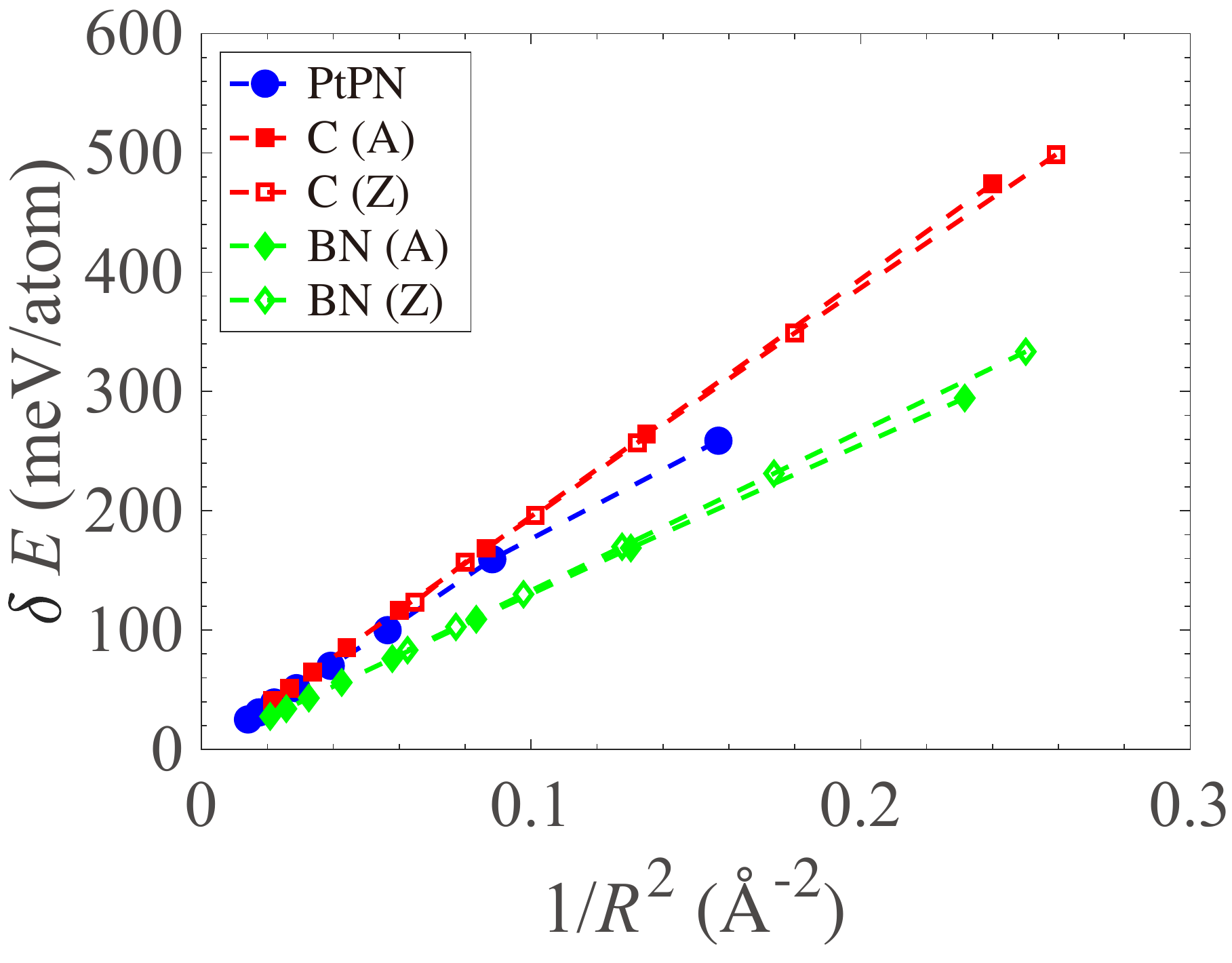}
  \caption{Variation of energy difference $\delta E$ of PtPN nanotubes with reference to their corresponding 2D sheets with $1/R^{-2}$, where $R$ denotes the radii of the nanotubes. A and Z in the brackets stand for armchair and zigzag, respectively.}
  \label{fig:fr}
\end{figure}

To quantify the feasibility of obtaining PtPN nanotubes from single-layer PtPN sheets, we convert $N$ to $R$ and adopt the following model describing the relationship between $\delta E$ and $R^{-2}$:\cite{landau1986theory, kudin2001c}
\begin{equation}
\delta E = \frac{D}{2} R^{-2},
\label{eq0}
\end{equation}
where $D$ is called the flexural rigidity also known as the bending stiffness of nanotubes. $D$ is a metric of the requirement of a force couple to bend the nanosheet per unit curvature.\cite{ru2000effective} The flexural rigidity of single-layer PtPN nanosheet arises from the combined effects of the resistance from both in-plane bond angle changes and out-of-plane electron clouds overlapping from Pt, P, and N atoms.\cite{pantano2004mechanics} Figure~\ref{fig:fr} shows the variation of $\delta E$ with $R^{-2}$ for PtPN nanotubes. In our recent work, we calculated $\delta E$ for armchair and zigzag C and BN nanotubes obtained from bending their nanosheets.\cite{liu2019dimension} We therefore plot the same variations for C and BN nanotubes shown in Fig.~\ref{fig:fr}. By linear fitting the $\delta E$ and $R^{-2}$ data to Eq.~\ref{eq0}, we compute and list $D$ for the different systems in Table~\ref{summary2}. The $D$ results of both C and BN nanotubes are consistent with previous work.\cite{kudin2001c} We also observe that the $D$ values for PtPN, C, BN nanotubes are comparable and the flexural rigidity of the PtPN nanotubes lies between those of C and BN nanotubes, which have both been available in experiment,\cite{janas2018towards} indicating that it is also feasible to obtain PtPN nanotubes.

\begin{figure}
  \includegraphics[width=8cm]{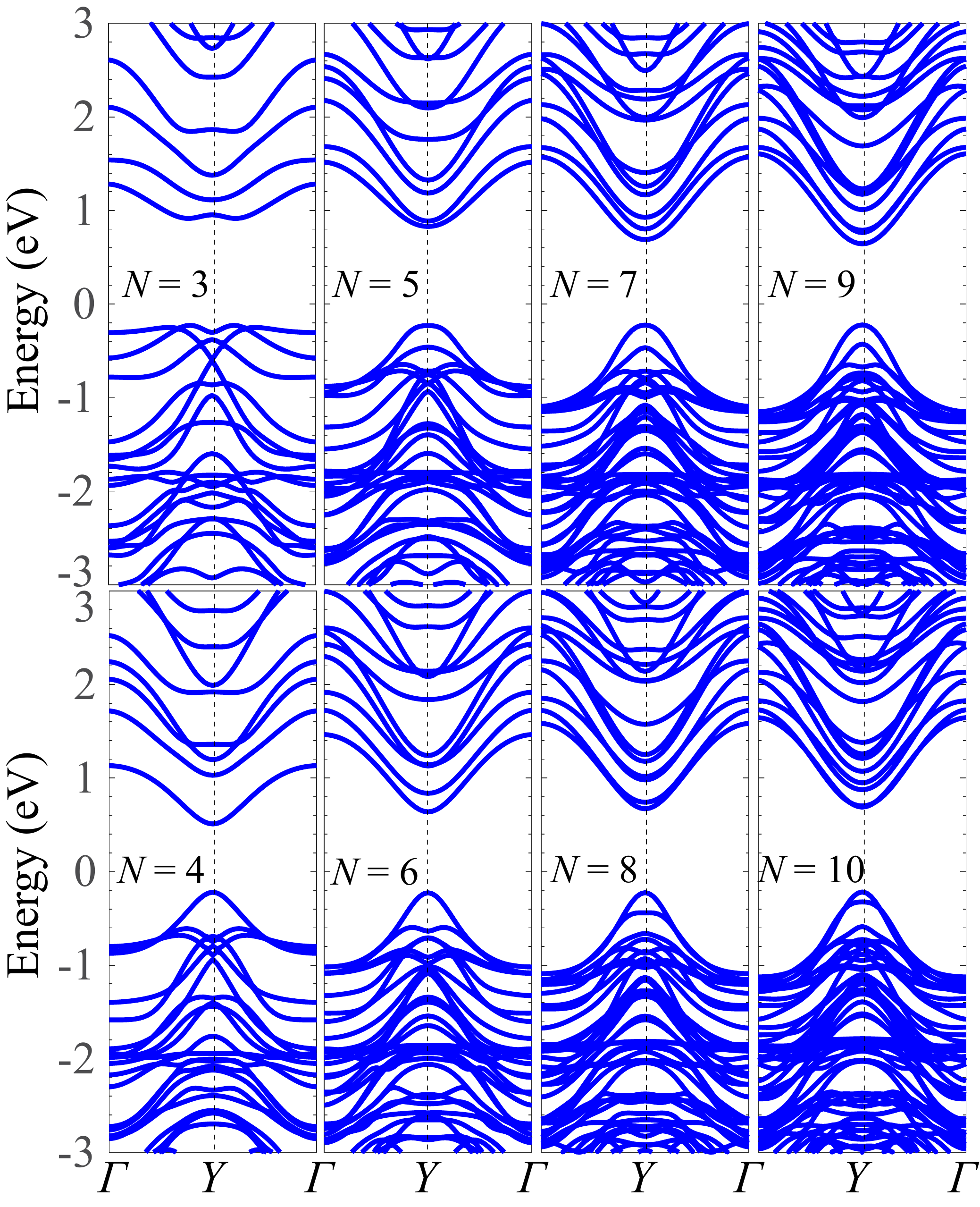}
  \caption{PBE band structures for PtPN nanotubes formed from $N \times 1 \times 1$ supercells of single-layer PtPN.}
  \label{fig:ntbands}
\end{figure}

\begin{figure}
  \includegraphics[width=8cm]{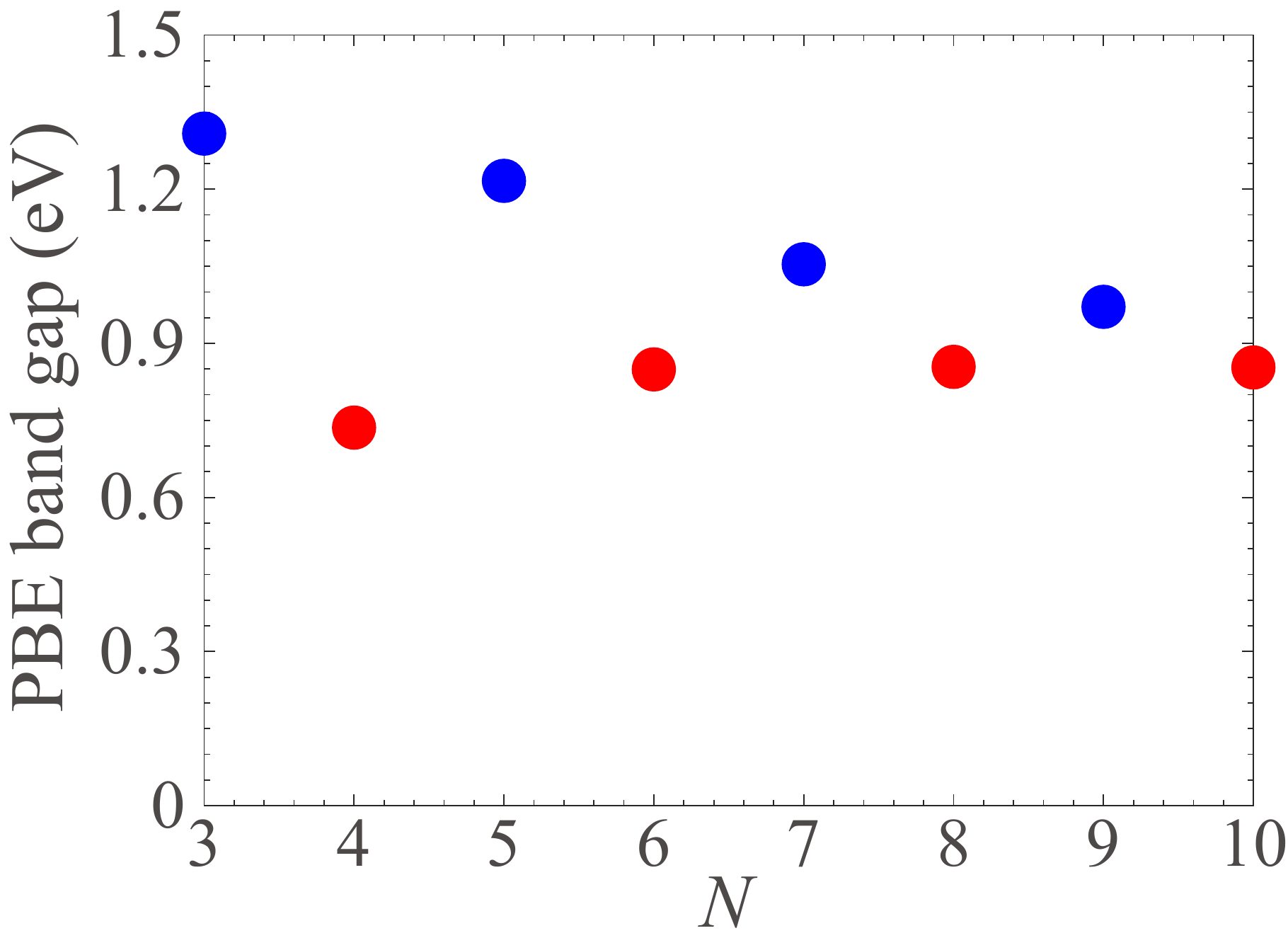}
  \caption{Dependence of PBE band gaps of PtPN nanotubes on $N$, which determines the number of unit cells of single-layer PtPN that form a nanotube.}
  \label{fig:bgs}
\end{figure}

\begin{figure}
  \includegraphics[width=8cm]{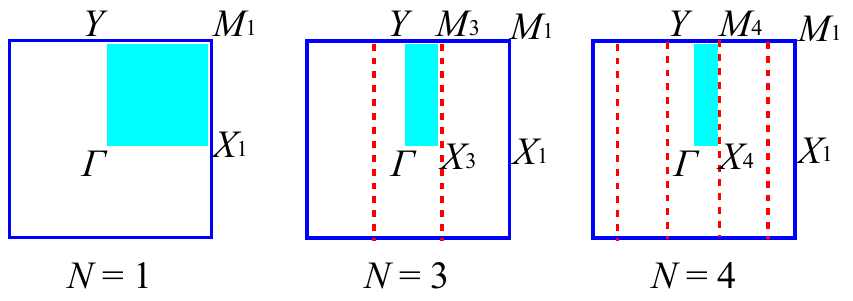}
  \caption{A quarter of the first Brillouin zones (represented by the cyan shaded areas) of $N \times 1 \times 1$ ($N$ = 1, 3, and 4) supercells of single-layer PtPN. The red dashed lines divide the first Brillouin zone (enclosed by the solid blue lines) of a unit cell of single-layer PtPN into $N$ equal portions.}
  \label{fig:triangle}
\end{figure}

\begin{table}
\caption{Predicted flexural rigidity $D$ (in eV$\cdot${\AA}$^{2}$/atom) of C, BN, and PtPN nanotubes. A and Z in the brackets represent armchair and zigzag, respectively.}
\begin{ruledtabular}
\begin{center}
\begin{tabular}{ccccc}
 C (A) & C (Z) & BN (A) & BN (Z) & PtPN\\
\hline
3.96 & 3.85 & 2.54 & 2.67 & 3.31\\
\end{tabular}
\end{center}
\end{ruledtabular}
\label{summary2}
\end{table}

Finally, we calculate the PBE band structures of the eight PtPN nanotubes shown in Fig.~\ref{fig:ntbands}, revealing that all of these nanotubes are direct-gap semiconductors. The variation of the band gaps of the PtPN nanotubes with $N$ is shown in Fig.~\ref{fig:bgs}. We observe that as $N$ increases, the band gaps of PtPN nanotubes with odd and even $N$ values decrease and increase, respectively, and appear to converge to a constant ($\sim$ 0.9 eV, close to the PBE band gap, 0.84 eV,  of single-layer PtPN) if $N$ is beyond 10. A similar dependence of band gaps on $N$ is also found in NiP$_2$ nanotubes.\cite{qian2019penta} Furthermore, the band gaps of the PtPN nanotubes with odd $N$ are wider than the nanotubes with even $N$. The relationship between the band gaps and $N$ shows that controlling the radii of PtPN nanotubes can tune their band gaps. 

To understand the larger band gaps of PtPN nanotubes when $N$ is odd, Fig.~\ref{fig:triangle} shows the high-symmetry $k$ points $\Gamma$, $X_N$, $M_N$, and $Y$ in the first Brillouin zones of single-layer PtPN represented by $N \times 1 \times 1$ ($N$ = 1, 3, and 4) supercells. For $N$ = 1, the $X_1$ and $M_1$ points are the same as the $X$ and $M$ points, respectively, as denoted in Fig.~\ref{fig:bandstructure}. The band gap at the $M$ point is smaller than that at the $X$ point calculated with either the PBE or HSE06 functional. When $N$ is larger and odd, e.g., $N$ = 3, the first Brillouin zone shrinks by three times, and the wave vectors and their corresponding energy levels along the $X_1$-$M_1$ direction are zone-folded to the $X_3$-$M_3$ direction, different from the $\Gamma$-$Y$ direction that is common for any $N \times 1 \times 1$ supercell. By contrast, if $N$ is even, e.g., $N$ = 4, the energy levels for the wave vectors along the $X_1$-$M_1$ direction will overlap with the energy levels of the wave vectors along the $\Gamma$-$Y$ direction. As a result, we can observe the band gap (originally at the $M$ point) along the $\Gamma$-$Y$ direction. For PtPN nanotubes, only the wave vectors along the $\Gamma$-$Y$ direction are allowed, so the even and odd $N$ lead to the occurrence and absence of the overlap along the $\Gamma$-$Y$ direction, respecively. The band gaps of PtPN nanotubes with odd $N$ are therefore wider than those of PtPN nanotubes with even $N$.

\section{Conclusions}
In summary, we predict a single-layer alloy PtPN with DFT calculations. This novel single-layer material consists of a pentagonal pattern and is completely planar and dynamically stable. We also find that single-layer PtPN exhibits direct band gaps of 0.84 and 1.60 eV calculated with the PBE and HSE06 functionals, respectively. Given the generally more accurate band gaps described by a hybrid density functional, the HSE06 band gap ensures a variety of promising optoelectronics applications of single-layer PtPN. We suggest that single-layer PtPN can be obtained from exfoliating bulk PtPN with a relatively low energy and the bulk pyrite-type structure can be acquired from alloying bulk PtP$_2$ with N atoms. We finally show that bending single-layer PtPN into the nanotube form result in nanotubes that exhibit tunable band gaps dependent on the radii of the nanotubes.
\begin{acknowledgments}
We thank the start-up funds from Arizona State University. This research used the computational resources of the AGAVE computer cluster at Arizona State University. 
\end{acknowledgments}

\end{document}